\pdfoutput=1
\documentclass[preprint,12pt,authoryear]{elsarticle}

\usepackage[T1]{fontenc}
\usepackage[utf8]{inputenc}
\usepackage{amsmath,amssymb}
\usepackage{booktabs}
\usepackage{tabularx}
\usepackage{graphicx}
\usepackage{array}
\usepackage{multirow}
\usepackage{float}
\usepackage{xcolor}
\usepackage{setspace}
\usepackage{caption}
\usepackage[expansion=false]{microtype}
\usepackage{xurl}
\graphicspath{{./}}

\newcommand{\toplines}[1]{\begin{tabular}[t]{@{}l@{}}#1\end{tabular}}
\usepackage[colorlinks=true,linkcolor=blue,citecolor=blue,urlcolor=blue]{hyperref}

\journal{Journal of Information Security and Applications}

\begin{document}

\begin{frontmatter}

\title{Auditing Anonymous AI Models: A Four-Stage Protocol for Black-Box Identity Verification}

\author{Yisen Xi}
\ead{xys21@tsinghua.org.cn}
\address{Independent Researcher, Beijing, China}

\begin{abstract}
The 2025--2026 AI market has seen a wave of \emph{stealth releases}: frontier models launched anonymously on developer platforms under codenames. For their users, identity determines data-handling terms, supply-chain risk, and capability expectations. No validated methodology exists for black-box identity verification of anonymous models: practitioner checklists lack accuracy evidence, and self-identification is untrustworthy by design. We propose a four-stage forensic audit protocol for API-served models. Stage 0 reconstructs launch-time configuration from archived platform snapshots (Internet Archive), exposing preview--production drift. Stage 1 fingerprints configuration (context, output ceiling, reasoning, modality) against the platform catalog. Stage 2 tests tokenizer identity with a cross-length differential that rejects short-prompt collisions. Stage 3 corroborates with behavioral probes. We test \emph{declaration consistency} on 10 known-identity releases (7 exact, 2 precision-differences, 1 partial, 0 counter-directional), not end-to-end identification under anonymity. Identification is validated prospectively on a flagship case whose 2026-08-23 analysis pointed to the GLM-5.3 version line and whose official reveal confirmed those family and version-line inferences (deployment variant was not pre-asserted; Flash was consistent post-reveal), and on three Stage-0-only cases where the protocol produced a graded hypothesis or declined rather than guessed. A standard-library-only implementation is provided as supplementary material.
\end{abstract}


\end{frontmatter}

\section{Introduction}\label{introduction}

\subsection{The problem}\label{the-problem}

On August 20, 2026, the reasoning model \emph{Ox Alpha} appeared on the OpenRouter API platform: a 1,048,576-token context window, multimodal input, tool calling, price zero, and no vendor attribution: a ``stealth model\ldots{} developed and operated by a third-party provider who has chosen to remain anonymous.'' Such anonymity is not incidental: anonymous models are \emph{engineered} not to reveal themselves (approximately 300 community identity-attack probes elicited zero credible self-identifications), and platform infrastructure obfuscates usage data. Yet identity is operationally decisive for users: data-retention terms are channel- and provider-specific, supply-chain risk (who serves the model? on what infrastructure?) is uninsurable without attribution, and capability expectations (what family? what training era?) shape engineering decisions.

This is not a single-case curiosity. Between 2025 and 2026 we enumerate 20 stealth-release cases from nine vendors across three distribution channels (built from public platform records and archived snapshots; full list in Supplementary Material S1); of these, 10 have both a launch-time specification record and a known identity and form the retrospective sample in Section 4. Anonymous models have appeared among high-usage entries on developer platforms. The verification problem (\emph{given an anonymous model, determine its true identity}) is therefore a digital-forensics and serving-integrity task: reconstruct identity from artifacts the serving stack cannot easily fake, and produce an auditable evidence trail for adoption and compliance decisions. Independent audits of \emph{claimed-official} third-party endpoints likewise find widespread identity mismatch \citep{zhang2026b}: 45.83\% of fingerprint checks failed on shadow APIs. This is evidence that API model identity cannot be taken on trust even when a brand is advertised. It also connects to established security concerns: model-extraction attacks showed that API access leaks model functionality \citep{tramer2016}, supply-chain attacks on open-source dependencies are a documented, growing threat class \citep{ohm2020}, and frontier-model identity determines the trust basis of every downstream integration \citep{openai2023}.

\subsection{Why not ask the model? Why not trust disclosures?}\label{why-not-ask-the-model-why-not-trust-disclosures}

Self-identification is a contaminated signal: anonymous models are system-engineered to self-present only under their codename, and community tests show both successful suppression (zero self-confessions under adversarial prompting) and unreliable self-claims (local deployments of the same family misidentifying as Claude). Vendor and platform disclosures are authoritative \emph{when they occur}, but they are the output of an adoption process that may never happen (one corpus case has remained anonymous for over eight months), and disclosures are not auditable. Practitioner protocols \citep{reed2026} provide a disclosure-evidence ladder and reproducible test protocol but publish no accuracy evidence, no retrospective validation, and no failure-mode analysis.

\subsection{Contributions}\label{contributions}

We contribute: (1) a \textbf{four-stage forensic audit protocol} for black-box identity verification of anonymous models, including a Stage 0 archived-configuration forensics method (time-consistent specification baseline via Internet Archive / record extracts) and a Stage 2 cross-length tokenizer differential test with a documented collision failure mode; (2) \textbf{retrospective declaration-consistency validation} on 10 stealth releases with known identities (7 exact, 2 precision-differences, 1 partial, 0 counter-directional); (3) \textbf{prospective blind validation} on the flagship Ox Alpha case, whose 2026-08-23 dated measurement artifacts supported a Zhipu GLM-5.3 version-line attribution later confirmed by the official reveal (ZAI GLM-5.3-Flash, 2026-08-26); deployment variant was not pre-asserted, plus a two-outcome tracking design for a long-running anonymous model (Big Pickle); (4) \textbf{negative controls} (known-backend re-acceptance; non-family collision rejection) and \textbf{``cannot-attribute'' demonstrations}; (5) a \textbf{standard-library-only implementation} (audit-tool) plus an auditor worksheet template, submitted as supplementary material; and (6) \textbf{scoped portability} to branded-base and serving-switch audits, distinguished from unvalidated end-to-end product studies.

\subsection{Scope and disclosure}\label{scope-and-disclosure}

The protocol is for settings where vendor or platform disclosure is absent, delayed, or disputed; it does not replace disclosure when one exists. All experiments used public API endpoints with zero-cost or vendor-approved access and respect platform terms. Findings are labeled ``identity unconfirmed, reproducibility limited'' where applicable (per provider data-retention terms). The flagship case was re-verified against the known model after official reveal (Section 5.1: constant tokenizer differential of 75 tokens across all four paired probes, four days post-reveal).

\section{Background and Related Work}\label{background-and-related-work}

\subsection{Model identification in AI systems}\label{model-identification-in-ai-systems}

Prior work on model identification falls into four streams.

\emph{AI-generated content detection} \citep{gehrmann2019,mitchell2023,bao2024} attributes authorship of text, not model provenance.

\emph{Watermarking and output fingerprinting} \citep{kirchenbauer2023,szyller2021,hu2024} require model-side cooperation or pre-existing fingerprints, absent in anonymous previews. Related work on IP protection fingerprints \citep{xu2025}, adversarial robustness of LLM fingerprints \citep{nasery2025}, and robustness of fingerprint-style inference \citep{szyller2022} shares the reliability concern but still targets cooperative settings.

\emph{Active black-box fingerprinting} sends crafted queries and classifies responses. LLMmap \citep{pasquini2025} combines expertise-informed probes with a classifier over the LLM version behind an application. ZeroPrint \citep{shao2026,shao2025} estimates output-distribution gradients and provides LeaFBench for copyright auditing. \citet{bruckner2026} verifies a \emph{claimed} OpenRouter endpoint against an enrolled reference using single-token answer distributions. FLIPS \citep{richardeau2026} targets instance-level IDs via pseudo-random sequences at finer granularity than family attribution.

These fingerprinting lines assume a cooperative or non-resistant target: response style and output statistics that stealth models are engineered to suppress (Section 1.1). They also omit prospective blind validation, explicit decline outputs, and launch-time baselines. We therefore rank deployment properties the target cannot easily fake: archived declarations (Stage 0), catalog fields (Stage 1), tokenizer differentials (Stage 2) over behavioral corroboration (Stage 3).

\emph{Gateway substitution audits} test whether the served backend matches what was advertised. Prior membership-inference and extraction work shows that API access leaks model information \citep{shokri2017,jagielski2020,tramer2016,carlini2024}. IRIS \citep{zhang2026a} detects stream substitution and routing dilution; Stability Monitor \citep{leshin2026} flags serving changes from output-distribution drift. A concurrent shadow-API audit \citep{zhang2026b} reports that 45.83\% of audited endpoints advertising frontier brands failed LLMmap fingerprint checks, with large utility gaps even when branding matched.

Those audits answer \emph{whether} a declared backend is faithful or \emph{when} it changed. Our setting is different: \emph{who} serves when no truthful declaration exists. Cross-provider serving deviations nonetheless overlap with our Stage 1/2 signals.

Across these strands, none validates on deliberately anonymized models, none pairs a declaration-consistency sample with a prospective blind test, and none treats \emph{cannot attribute} as a first-class output. Table 1 summarizes the gap.

\subsection{Tokenizer-based attribution}\label{tokenizer-based-attribution}

Tokenizer fingerprints have been used for model-family attribution in community practice (e.g., counting prompt tokens across candidate models to detect shared tokenizers). Two limitations are known but under-documented: (i) short prompts can produce \emph{coincidental} equal token counts across different tokenizers (we document empirical counterexamples: three non-candidate families exhibited constant differentials on short prompts that collapsed under long-text testing); and (ii) gateway-level usage reporting can be transformed by the intermediary, so raw token counts require differential (between-model) rather than absolute interpretation. Our cross-length differential test (Section 3.3) addresses both limitations.

\subsection{Practitioner protocols and community forensics}\label{practitioner-protocols-and-community-forensics}

\citet{reed2026} proposes a disclosure-evidence ladder for stealth models: evidence channels (platform disclosure, provider behavior, model card, etc.) graded as \emph{confirmed}, \emph{inferred}, or \emph{unknown}, plus a reproducible test protocol over technical and behavioral signals. The protocol has no retrospective validation, no failure-mode analysis, and relies on disclosures rather than active measurement. \citet{shrey2026} conducted a large community forensics campaign on a single flagship case (600+ requests, 13.5M prompt tokens): 44-string tokenizer differentials, knowledge-boundary tests, context-limit sweeps, and modality tests, concluding GLM-5 generation attribution. The constant tokenizer-count offset on that case was first reported there; our contribution is the cross-length constancy criterion, collision rejection, and protocolization, not the original observation. That campaign validated the power of tokenizer differentials at scale but was case-specific, ad hoc, and, as the official reveal showed, less precise than configuration-layer evidence at the version level (the campaign's knowledge-boundary inference pointed to the base checkpoint; configuration evidence pointed to the version line that the reveal confirmed). Our protocol systematizes these dispersed techniques into a staged, validated, tooled procedure, and adds the Stage 0 time-consistent baseline that the community campaign lacked.

\subsection{Positioning}\label{positioning}

Relative to the systems in Section 2.1--2.3, we differ mainly in four places: a declaration-consistency sample (retrospective N=10), prospective blind testing (including a confirmed flagship case), an explicit \emph{cannot attribute} output, and a standard-library-only implementation. Table 1 situates these gaps against the closest related systems (signals and mechanisms are discussed in Section 2.1--2.3). Scope relative to vendor disclosure is stated in Section 1.4.

\begin{table}[H]
\centering
\caption{Comparison with related identification / serving-integrity methods.}
\label{tab:1}
\footnotesize
\setlength{\tabcolsep}{4pt}
\begin{tabularx}{\textwidth}{@{}XXXXX@{}}
\toprule
Method & Setting & Coop. & Prospective & Decline \\
\midrule
\citet{reed2026} & Stealth releases & Partial & No & Graded labels only \\
LLMmap \citep{pasquini2025} & Known models behind apps & Yes & No & Open-set $\neq$ decline \\
ZeroPrint \citep{shao2025,shao2026} & Fingerprint / copyright & Yes & No & No \\
\citet{bruckner2026} & Claimed OpenRouter endpoints & Yes & No & Vs enrolled reference \\
FLIPS \citep{richardeau2026} & Instance / config ID & Yes & No & Classifier only \\
IRIS \citep{zhang2026a} & Gateway substitution & Yes & No & Deviation flag \\
Shadow-API audit \citep{zhang2026b} & Claimed third-party APIs & Yes & No & Mismatch vs official \\
Stability Monitor \citep{leshin2026} & Serving change detection & Yes & Change-points only & Change $\neq$ identity \\
\textbf{This work} & \toplines{\textbf{Anonymous}\\\textbf{APIs}} & \toplines{\textbf{No}\\\textbf{(Stages 0--2)}} & \toplines{\textbf{Yes}\\\textbf{(Ox Alpha)}} & \textbf{Yes} \\
\bottomrule
\end{tabularx}
\end{table}

\emph{Note.} \emph{Coop.} = needs cooperative behavioral or distributional responses. \emph{Prospective} = anonymous pre-reveal test later scored against an official reveal. \emph{Decline} = first-class \emph{cannot attribute} (or equivalent audit stop) rather than forced classification.

\section{The Audit Protocol}\label{the-audit-protocol}

The protocol has four stages (Stages 0--3), with a mandatory cross-length guard inside Stage 2 (Fig. 1). It is implemented in audit-tool (zero third-party dependencies; keys read from environment, never logged or stored). The intended user is an IT-security or supply-chain auditor who must decide whether to adopt an anonymous endpoint before a vendor disclosure exists; each stage produces artifacts intended to form an auditable evidence chain consistent with forensic incident-response practice \citep{nist2006}. For prospective claims, auditors should \emph{commit} graded conclusions (e.g., SHA-256 of a prediction card) to a public timestamped channel before reveal; dated measurement logs alone are weaker under adversarial scrutiny of self-held clocks.

\begin{figure}[H]
\centering
\includegraphics[width=\textwidth,keepaspectratio]{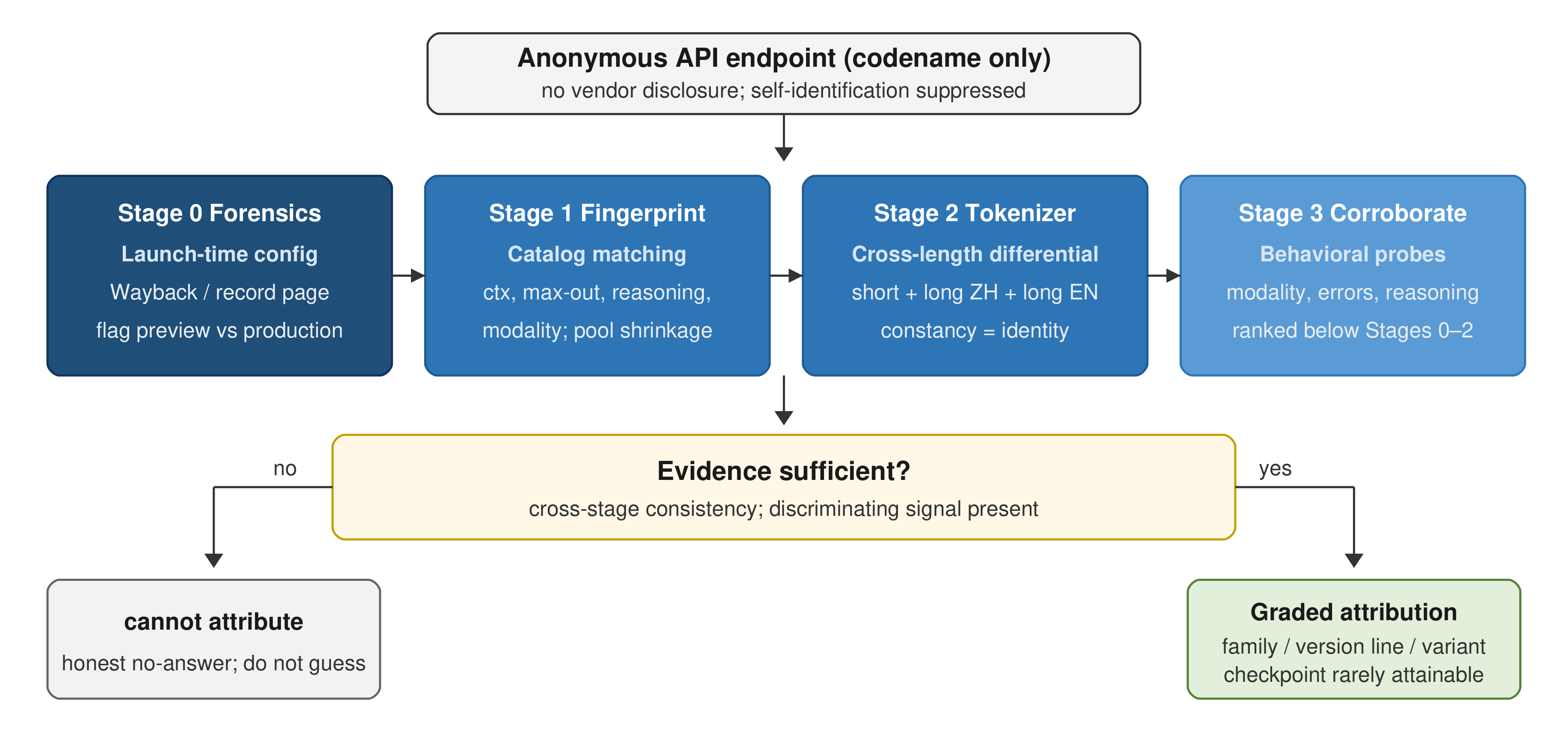}
\caption{Four-stage forensic protocol for anonymous-model identity verification. Stage 0 reconstructs the launch-time configuration from archived platform records. Stage 1 shrinks the live catalog to a configuration-compatible candidate pool. Stage 2 tests tokenizer identity with a mandatory cross-length differential (short-prompt constancy is not evidence). Stage 3 corroborates with behavior. Insufficient evidence yields \emph{cannot attribute} rather than a forced guess.}
\label{fig:1}
\end{figure}

\subsection{Stage 0: Configuration-declaration forensics (time-consistent baseline)}\label{stage-0-configuration-declaration-forensics-time-consistent-baseline}

\emph{Goal.} Establish what the target \emph{declared} at launch time, and whether the declaration drifted over time.

\emph{Procedure.} (1) Locate archived snapshots of the model's platform record page (Internet Archive; also platform /models endpoints) at launch and at regular intervals; coverage of those pages is the critical dependency here \citep{ainsworth2011}. (2) Extract configuration fields: context window, max output, reasoning controls, modality (text/image/video/audio), supported parameters. (3) Compare launch-time vs.~current declarations; flag drift (we document version drift between preview and production configurations in a flagship case: preview 1,048,576 $\rightarrow$ production 1,310,720 context, with modality constant, see Section 5). (4) Record the \emph{launch-time} configuration as the comparison baseline for Stages 1--3.

\emph{Rationale.} Configuration is a deployment property, not a model-weights property; it varies across snapshots and channels. Using current specifications to test a model whose preview configuration differed invites false negatives (we document this as a retrospective method-correction lesson, Section 4.1). Stage 0 converts the ``time-point scanning'' principle of disclosure auditing into operational forensics. Table 2 records the two central drift cases used in this study; the full extraction sheet is Supplementary Material S3. Auditors are expected to produce a portable worksheet (Supplementary Material S2) rather than an informal notebook.

\begin{table}[H]
\centering
\caption{Stage 0 drift cases (launch-time vs later declaration; full extracts in S3).}
\label{tab:2}
\footnotesize
\setlength{\tabcolsep}{4pt}
\begin{tabularx}{\textwidth}{@{}XXXX@{}}
\toprule
Case & Launch ctx & Later / official & Role \\
\midrule
Ox Alpha (preview) & 1,048,576 & 1,310,720 (GLM-5.3-Flash) & Prospective baseline \\
Polaris Alpha & 262,144 & 400K (GPT-5.1) & Retrospective partial / drift lesson \\
\bottomrule
\end{tabularx}
\end{table}

\subsection{Stage 1: Configuration fingerprinting against the catalog}\label{stage-1-configuration-fingerprinting-against-the-catalog}

\emph{Goal.} Shrink the candidate pool from the platform's full model catalog to a small set of configuration-compatible candidates.

\emph{Procedure.} (1) Pull the platform's full model catalog (e.g., OpenRouter /api/v1/models). (2) Build the target's configuration fingerprint from the launch-time record (Stage 0): context window, max output, reasoning controls, modality, supported parameters. (3) Match the fingerprint against all catalog entries; report the candidate set \textbf{as a layered shrinkage, not a single pool}: matching on context window and max output alone (the two most common fields) shrank the flagship-case catalog from 422 entries to 6 (a 70$\times$ reduction); adding modality produced a \emph{unique} match. The conservative pool (context + max output) is the candidate set carried into Stage 2, since it preserves recall over vendor families; the unique modality-level match is recorded as a strong auxiliary signal with a caveat: in the flagship case, the anonymous target declared video input while its true public counterpart did not; post-hoc, this was an \emph{early leakage of the variant's multimodal capability} ahead of the public catalog's declaration, but at audit time it was indistinguishable from declaration noise. Stage 1 therefore reports both pools and flags modality divergence between target and candidates rather than discarding candidates on it. (4) Record modality separately as a \emph{scarce time-invariant signal}: modality is less subject to context inflation and version drift than context length (retrospective evidence: modality identified a variant that context length could not), while its \emph{declaration} can lag serving capability, making Stage 0's launch-time baseline the arbiter of which declaration is authoritative.

\subsection{Stage 2: Cross-length tokenizer differential test (mandatory guard)}\label{stage-2-cross-length-tokenizer-differential-test-mandatory-guard}

\emph{Goal.} Establish tokenizer identity (same tokenizer implies same family/generation) with a false-positive-resistant test.

\emph{Procedure.} (1) For the target and each candidate, query the same prompt and record \emph{prompt\_tokens} from usage fields; compute the difference. (2) Repeat across $\geq$3 texts spanning short (identity questions), long Chinese, and long English texts (\textbf{mandatory: at least one long text per language family}). (3) \textbf{Constancy rule ($\varepsilon$ = 0):} the target and candidate are treated as sharing a tokenizer only if the differential is \emph{exactly equal} across all \emph{retained} probes (same tokenizer implies the difference equals the fixed hidden-system-prompt injection delta, which is prompt-length invariant). Near-matches are not accepted. (4) \textbf{Probe-stability rule (protocol-level, applied before the constancy test):} for each probe id, if the \emph{target} endpoint returns inconsistent \emph{prompt\_tokens} across repeated identical rounds, discard that probe from the constancy assessment for all candidates; do not tune the probe set after seeing family outcomes. (5) Counterexample rule: if a constant differential appears on short prompts only and collapses on long texts, classify as \emph{collision} (different tokenizers coincidentally agreeing on short token counts), not evidence.

\emph{Failure-mode documentation.} We empirically document three collision cases on \emph{non-family} endpoints (step-3.7-flash, hy3, kimi-k3): on short identity prompts they exhibited constant (or near-constant) differentials (75, 72, and 2 tokens respectively) that dissolved under the archived 20-probe ox-full battery (Fig. 2 source JSONL): step-3.7-flash ranged over 69--77, hy3 over 68--77, and kimi-k3 over --1 to 5, while the true family held a constant 75 across every paired probe of that battery and across the five canonical cross-length probes (short$\times$3 + long Chinese + long English). A separate 16-probe archival run shows the same family constant at 75 on 15/16 probes; applying the probe-stability rule above, a style probe (\emph{style\_en}) was discarded because the \emph{target} alone was round-unstable (105 then 20 tokens) while candidates remained stable at 30) before computing constancy. Target-side instability is itself a serving-path signal (template, caching, or non-deterministic accounting), not evidence for or against a candidate family. Without the mandatory multi-length condition, the short-prompt collisions would have produced false positives. Same-generation siblings that \emph{do} share a tokenizer (in the flagship run, glm-5.2 with glm-5.3) remain constant across lengths; that is family-level identity, not a collision, and version-line separation is left to Stage 1. Hence \textbf{cross-length constancy is the criterion; short-prompt constancy alone is not evidence}. For the short-prompt-only comparison in Table 3 we use a near-constant acceptance rule (differential spread $\leq$ 2 tokens across the three identity/cutoff probes counts as constant), reflecting the coarser quantization of short prompts; under this rule kimi-k3's $\Delta$ $\in$ \{2, 4\} pattern is admitted as a collision, and the short-only rule yields two true acceptances plus three false positives.

\emph{Collision ablation.} The mandatory multi-length condition is the central component of Stage 2, not a stylistic precaution. Replaying the flagship-case ox-full run under two decision rules (Fig. 2; Table 3) shows that short-prompt-only constancy admits three non-family false positives, while the full multi-length rule admits only the true family. Multi-length structure is what converts a heuristic into a criterion; with a single short string, collision and identity are indistinguishable.

\begin{figure}[H]
\centering
\includegraphics[width=\textwidth,keepaspectratio]{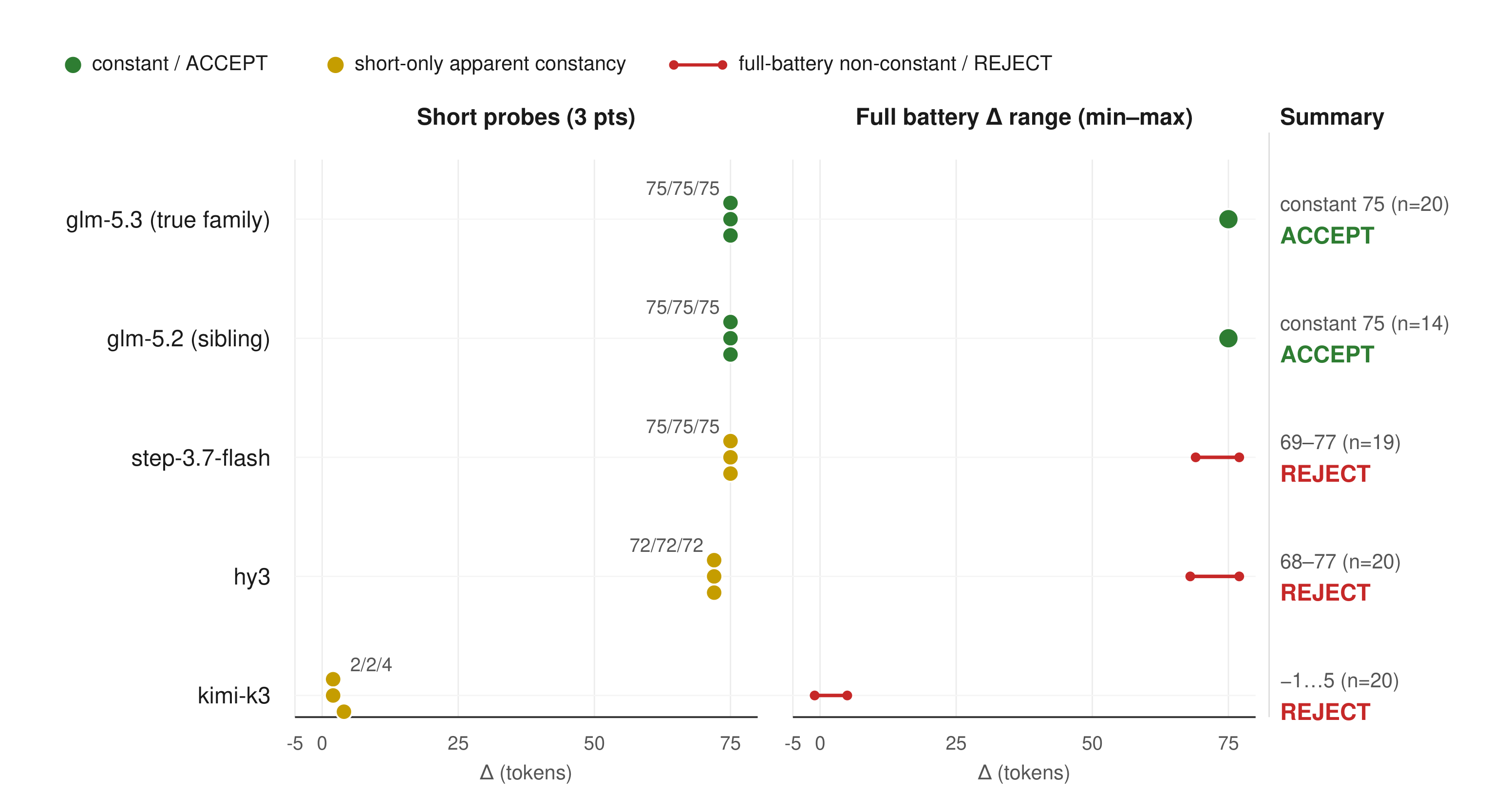}
\caption{Tokenizer differentials (ox $\rightarrow$ candidate) from the archived ox-full JSONL (2026-08-23); $\Delta$ = prompt\_tokens(ox) -- prompt\_tokens(candidate). Left: three short-probe $\Delta$ points; right: full-battery $\Delta$ range (min--max) over paired probes.}
\label{fig:2}
\end{figure}

\begin{table}[H]
\centering
\caption{Stage 2 outcomes on the five Fig. 2 endpoints.}
\label{tab:3}
\footnotesize
\setlength{\tabcolsep}{8pt}
\begin{tabularx}{\textwidth}{@{}Xrrrr@{}}
\toprule
Decision rule & TP & FP & TN & FN \\
\midrule
Short-prompt only & 2 & 3 & 0 & 0 \\
Full multi-length ($\varepsilon$ = 0) & 2 & 0 & 3 & 0 \\
\bottomrule
\end{tabularx}
\end{table}

\emph{Note.} Source: ox-full JSONL, 2026-08-23; true family = \{glm-5.2, glm-5.3\}. Short Accept includes kimi-k3's near-constant short pattern ($\Delta$ $\in$ \{2, 4\}); Full Accept uses $\varepsilon$ = 0 over paired ox-full probes.

Beyond the five Fig. 2 endpoints in Table 3, three further non-family endpoints in the same JSONL (ernie-, mimo-, and qwen-line) are rejected by the full multi-length rule (archive-wide: TP = 2, FP = 0, TN = 6, FN = 0); they also fail exact short $\varepsilon$ = 0 constancy and are outside the Fig. 2 collision panel.

\emph{Attribution boundary.} Tokenizer identity establishes \emph{family and generation} with high confidence; it does not establish exact checkpoint, quantization, or deployment variant (we document a case where tokenizer matched a generation while the true model was a new-architecture variant of that generation (same tokenizer family, different architecture; Section 5)). The differential design also neutralizes a gateway-level threat: intermediaries can transform absolute usage reports, so only \emph{between-model} differences on identical prompts are treated as evidence. We use SentencePiece-style subword segmentation \citep{sennrich2016,kudo2018} as the reference model of tokenizer behavior: subword segmentation is deterministic given the tokenizer and prompt, so cross-length constancy of the differential follows from tokenizer identity, and any deviation is diagnostic.

\emph{Formal basis.} The test's logic can be stated precisely. Let a serving endpoint \(E\) wrap a model with tokenizer \(T\) and prepend a fixed hidden system prompt \(s_E\) to every user prompt \(u\), reporting \(\mathrm{ptok}_E(u) = T(s_E \cdot u)\), where \(\mathrm{ptok}_E(u)\) denotes the prompt-token count reported by endpoint \(E\) for prompt \(u\), and \(\cdot\) denotes string concatenation. Under (A1) deterministic segmentation and (A2) faithful arithmetic reporting, and provided the prompt assembly is \emph{boundary-stable} (the template closes with a fixed delimiter so that segmentation of \(u\) is independent of the preceding context), \(T(s_E \cdot u) = T(s_E) + T(u)\). Consequently, for any two endpoints that share a tokenizer,

\[\mathrm{ptok}_{E_1}(u) - \mathrm{ptok}_{E_2}(u) = T(s_{E_1}) - T(s_{E_2}) \equiv \Delta \quad \text{for all } u,\]

That is, the differential is a prompt-length-invariant constant equal to the difference in hidden-injection lengths. The diagnostically useful direction is the \emph{contrapositive}: if two prompts \(u_1\) and \(u_2\) yield non-constant differentials (under the idealization \(\varepsilon = 0\), i.e., zero reporting noise), then at least one of (a) different tokenizers, (b) non-fixed injection, or (c) reporting transformation must hold; each of which independently disqualifies same-backend attribution. The converse (cross-length constancy implying the same tokenizer) is \emph{not} a general theorem; it is only a \emph{necessary} consequence of tokenizer identity under (A1)--(A2). We use it operationally because short-prompt collisions are common, whereas multi-length collisions are empirically rare in our archives (Fig. 2). Joint collision risk over the \(k\) length bands of the battery (\(k=5\) in the canonical cross-length probe set) need not factor into independent per-band probabilities (tokenizers often share BPE merge structure, so collisions can be positively correlated); multi-length rejection power is therefore treated as an \emph{empirical} operating characteristic rather than a multiplicative probability claim. Two boundary conditions further delimit the claim: a server-side template change alters \(s_E\) and thereby shifts \(\Delta\) (the version-drift event detected by Stage 0), and an (A2)-violating gateway can distort absolute counts (neutralized by the differential design). Empirically, cross-length constancy held for the true family through reveal (Section 5.1), and the documented short-prompt collisions were rejected by the multi-length rule (Table 3).

\subsection{Stage 3: Behavioral probes and corroboration}\label{stage-3-behavioral-probes-and-corroboration}

\emph{Goal.} Corroborate or discriminate candidates using behavior that candidates cannot easily fake: (1) \emph{knowledge boundaries} (model launch-time recall: knowing an earlier model but not the candidate generation's launch, noisy, and we rank it \emph{below} configuration and tokenizer evidence after the flagship case's reveal contradicted a knowledge-boundary inference; calibration research shows models partially encode what they do not know, \citet{kadavath2022}, but self-reported identity remains unreliable for deliberately anonymized models, Section 1.2); (2) \emph{upstream error codes and refusal patterns} (provider-specific HTTP error bodies, content-filter patterns); (3) \emph{reasoning controls} (/nothink behavior, reasoning-effort parameter handling); (4) \emph{modality capability tests} (image/video/audio handling; audio failure vs.~image success discriminated variant profiles in the flagship case); (5) \emph{multilingual behavior} (weak discriminator alone).

\subsection{Confidence grading and output discipline}\label{confidence-grading-and-output-discipline}

Attribution outputs carry a graded confidence: (i) \emph{family} (vendor/generation), (ii) \emph{version line}, (iii) \emph{variant} (flash/turbo/max), (iv) \emph{exact checkpoint} (rarely attainable). The protocol explicitly supports ``no-answer'' outputs: when evidence is insufficient (no discriminating signal, no configuration match, scarce signals uninformative at the time point), the correct output is \emph{cannot attribute}, not a forced guess. We demonstrate three Stage-0-only cases (Section 5.3) where the protocol produced a graded hypothesis or \emph{cannot attribute} rather than a forced identification; in one case, post-hoc verification confirmed the decline was correct.

\subsection{Threat model, detection boundaries, and portability}\label{threat-model-detection-boundaries-and-portability}

\textbf{Adversary classes.} The protocol is designed against four classes (classification adapted from standard threat-modeling practice, \citealp{shostack2014}):

\begin{itemize}
\item
  \textbf{A-idle (honest anonymous deployment)}: no adversary; the operator withholds identity but reports configuration truthfully. Baseline scenario: Stages 0--2 operate at full strength.
\item
  \textbf{A-declare (declaration manipulation)}: the operator falsifies the catalog record (inflated context, wrong modality) to mislead fingerprinting. Mitigations: Stage 0 anchors the \emph{launch-time} declaration from an independent archive (the attacker must have falsified the record at launch and maintained the forgery across snapshots), and several declared properties are \emph{behaviorally testable}: advertised context by boundary-probing, advertised modality by modality tests (Stage 3). Residual risk: undeclarable serving properties (quantization, inference stack) are untestable black-box, an acknowledged boundary.
\item
  \textbf{A-mimic (backend mimicry)}: a provider serves a different model behind the record of the advertised one (impersonation or undisclosed swap), the ML-supply-chain analogue of backdoored or swapped third-party components \citep{gu2017}. Detected when the served backend's configuration/tokenizer fingerprints diverge from the declaration or from reference samples. This is the division of labor with gateway-substitution auditing (Section 2.1): IRIS-class monitors detect \emph{that} a substitution occurred with statistical guarantees; our Stages 1--2 \emph{attribute which} model is now being served, without enrollment. Real-world shadow-API measurements illustrate a related boundary: fingerprint agreement with a claimed brand need not preserve task utility (\citealp{zhang2026b} report large MedQA drops for advertised Gemini endpoints): identity match is not behavioral fidelity, so Stage 3 remains a cross-check rather than a substitute for Stages 0--2.
\item
  \textbf{A-evasive (adversarial suppression)}: the operator engineers the model to resist identity leakage; the stealth-release case (Section 1.1: zero self-identifications under \textasciitilde300 community probes). Defeated by construction: Stages 0--2 consume deployment properties (declarations, catalog entries, usage arithmetic) that do not depend on model cooperation. Residual risk: a \emph{full-stack proxy} that forwards queries to a different model's API while rewriting usage fields; against this, the protocol offers a \textbf{cross-stage inconsistency alarm}: when Stage 2 attributes family X but Stage 3 behavior (self-identification content, vendor markers) is characteristic of family Y, the attribution is flagged as proxy-suspect. The alarm's \emph{positive} direction (stages agree) was observed post-reveal (Section 5.1). Its \emph{negative} direction is exercised in controlled Experiment E1 (Section 6.1; Supplementary Material S4): forging \emph{prompt\_tokens} to match a declared GLM endpoint while serving a Kimi backend makes Stage 2 accept the declared family (constant $\Delta$=0), while Stage 3 identity probes return Moonshot/Kimi self-reports, triggering the alarm in both offline JSONL replay and live OpenRouter runs.
\end{itemize}

\textbf{Detection boundaries.} Stage 0 trusts the archive's integrity (Wayback compromise is out of scope); Stage 2 trusts arithmetic reporting up to the differential design (A2 violations affecting both models symmetrically cancel; asymmetric forgery triggers the consistency alarm above); no stage detects inference-stack changes \emph{below token-count resolution}: quantization, kernel, or hardware swaps that preserve token counts are invisible to all three stages. That layer sits beside change-detection systems that fingerprint output distributions over time (Stability Monitor, \citealp{leshin2026}; they detect \emph{that} something changed; we attribute \emph{what it is}); together they cover a fuller serving-integrity audit.

\textbf{Portability.} Primary scenario: official anonymity (no declaration; inference from observable behavior). Secondary: shadow-API impersonation (a provider masquerading as a known model). Extended scenarios (Section 6): branded-model base attribution (declared model $\neq$ actual backbone, e.g., coding products built on third-party open backbones) and silent-downgrade detection (ad-funded free tools that declare a premium model but serve a cheaper variant under load, detectable because downgrade changes tokenizer/configuration fingerprints). The tool's configuration and tokenizer layers are reusable across these scenarios with the same probe sets.

\section{Retrospective Validation}\label{retrospective-validation}

\subsection{What the retrospective sample tests, and what it does not}\label{what-the-retrospective-sample-tests-and-what-it-does-not}

The 10 retrospective cases with known identities (six vendors: OpenAI, xAI, Xiaomi, Zhipu AI, NVIDIA, Mistral) test whether \emph{launch-time declared specifications (record pages) match the true models' official specifications}: that is, \textbf{declaration consistency and version drift}, the evidential foundation Stages 0--1 rest on. The sample is conditioned on having \emph{both} an archived launch-time specification and a later known identity (Supplementary Material S1); cases lacking either are out of scope for Table 4. That filter can favor semi-disclosed releases: selection on verifiability, not a random draw from all stealth endpoints, so Table 4 estimates channel reliability under observable records, not population identification accuracy. Because most corpus cases were semi-disclosed (the record page itself named the true model) or officially revealed, these cases do \emph{not} exercise the identification pipeline under anonymity: the protocol ran blind against unknown identity only in the prospective flagship case (Section 5.1), supported by the collision ablation and the head-to-head comparison (Sections 3.3, 4.4). We state this scope explicitly; the retrospective sample validates the \emph{evidence channel}, not end-to-end identification accuracy. The validation itself surfaced three method corrections, reported here as lessons:

\begin{enumerate}
\def\labelenumi{\arabic{enumi}.}
\item
  \textbf{v0.1 error: using current-listing models as comparison baseline} systematically underestimated match rates (1/7 exact) because later models served as the comparison; \textbf{v0.2 correction}: compare against the true identity's launch-time official specifications (third-party/official mixed), raising exact-match to 5/7.
\item
  \textbf{v0.3 correction (first-party official sources)}: re-verified all specifications against first-party official sources (vendor docs, arXiv) where available.
\item
  \textbf{v0.4 sample expansion (platform enumeration)}: systematic platform enumeration (Wayback; OpenRouter namespace; OpenCode Zen snapshots) added three cases (Andromeda Alpha, Bert-Nebulon Alpha, Pony Alpha record), N=7$\rightarrow$10, and corrected the earlier claim that Pony had no record page (it does: 200K context, semi-disclosed).
\end{enumerate}

\subsection{Results}\label{results}

Final retrospective outcomes (N=10, one codename = one case): \textbf{7 exact matches, 2 precision-differences, 1 partial match, 0 counter-directional} (Table 4).

\textbf{Verdict rubric (applied uniformly).} Comparisons are against the \emph{official declaration's semantics}, not a single numeric tolerance on raw integers. \textbf{Exact:} the record matches the vendor/platform official specification under that specification's own rounding or banding (e.g., an official ``1M'' context covers a 1,048,576-class declaration; ``1.8M'' covers 1,840,000). \textbf{Precision-difference:} both sides publish \emph{exact} integer windows that differ by \textless1\% with no official rounded band that absorbs the gap (Quasar/Optimus: 1,048,576 vs GPT-4.1's 1,047,576). \textbf{Partial:} identity direction is consistent, but a documented preview$\rightarrow$production (or snapshot$\rightarrow$card) drift changes a compared field (Polaris). \textbf{Counter-directional:} record points to a different vendor/family than the reveal (none in this sample).

\begin{table}[H]
\centering
\caption{Retrospective declaration-consistency sample (N=10).}
\label{tab:4}
\footnotesize
\setlength{\tabcolsep}{4pt}
\begin{tabularx}{\textwidth}{@{}XXXXX@{}}
\toprule
Codename & Record ctx & Official & Disclosure & Verdict \\
\midrule
Quasar Alpha & 1,048,576 & GPT-4.1 (1,047,576) & Platform & Precision (0.1\%) \\
Optimus Alpha & 1,048,576 & GPT-4.1 (1,047,576) & Platform & Precision (0.1\%) \\
Sonoma Dusk Alpha & 2,000,000 & Grok 4 Fast (2M) & Semi & Exact \\
Sherlock Think Alpha & 1,840,000 & Grok 4.1 Fast (1.8M) & Semi & Exact \\
Hunter Alpha & 1,048,576 & MiMo-V2-Pro (1M) & Semi+V & Exact \\
Healer Alpha & 262,144 + modality & MiMo-V2-Omni & Semi+V & Exact \\
Polaris Alpha & 262,144 & GPT-5.1 (400K) & Semi & Partial (drift) \\
Pony Alpha & 200K & GLM-5 (200K) & Semi+V & Exact \\
Andromeda Alpha & 128K & Nemotron Nano 2 VL & Semi & Exact \\
Bert-Nebulon Alpha & 256K & Mistral Large 3 & Semi & Exact \\
\bottomrule
\end{tabularx}
\end{table}

\emph{Note.} Verdicts follow the rubric above (not end-to-end identification under anonymity). \emph{Disclosure:} Platform = platform reveal; Semi = record semi-disclosure; Semi+V = semi-disclosure plus vendor claim.

Zero counter-directional cases: no record pointed toward a wrong vendor. A second same-launch codename, Sonoma Sky Alpha (same true model as case 3, launched the same day), is excluded because no independent record-specification snapshot was archived for it; including it as an eleventh case would double-count one specification match.

\subsection{Candidate-pool shrinkage (Stage 1 quantification)}\label{candidate-pool-shrinkage-stage-1-quantification}

For the \emph{prospective} flagship case (not one of the Table 4 rows), the Stage 1 matching procedure against the full catalog (N=422 models) was instrumented field-by-field (Fig. 3). The shrinkage funnel: context-window equality (1,048,576) retains 48/422 (11.4\%); adding max-output equality (131,072) retains 6 (1.4\%); adding the reasoning profile (mandatory reasoning, default effort = max) retains 3 (0.7\%); the supported-parameters set is identical for all 3; and adding modality (text+image+video$\rightarrow$text) yields a \textbf{single-model candidate set}, a 422$\rightarrow$1 reduction (99.8\% shrinkage) from configuration fields alone. Notably, the final filter that achieves uniqueness is modality, the signal we identify as scarce and time-invariant (F4): context and output ceilings are shared by several models, but the image+video input combination was unique in the catalog at that time point. The protocol therefore \emph{reports} L2 (size 6) as the Stage-2 candidate pool (recall-preserving) and L4 uniqueness as an auxiliary strong signal with the declaration-lag caveat of Section 3.2. Total query cost of the full audit (both protocol sides, all probes) is on the order of \(10^2\) API calls and \textasciitilde{}\(10^4\) prompt tokens, at the flagship's free-window pricing, \$0; at the confirmed model's list price, under a few cents. Attribution cost is thus dominated by latency, not spend.

\begin{figure}[H]
\centering
\includegraphics[width=\textwidth,keepaspectratio]{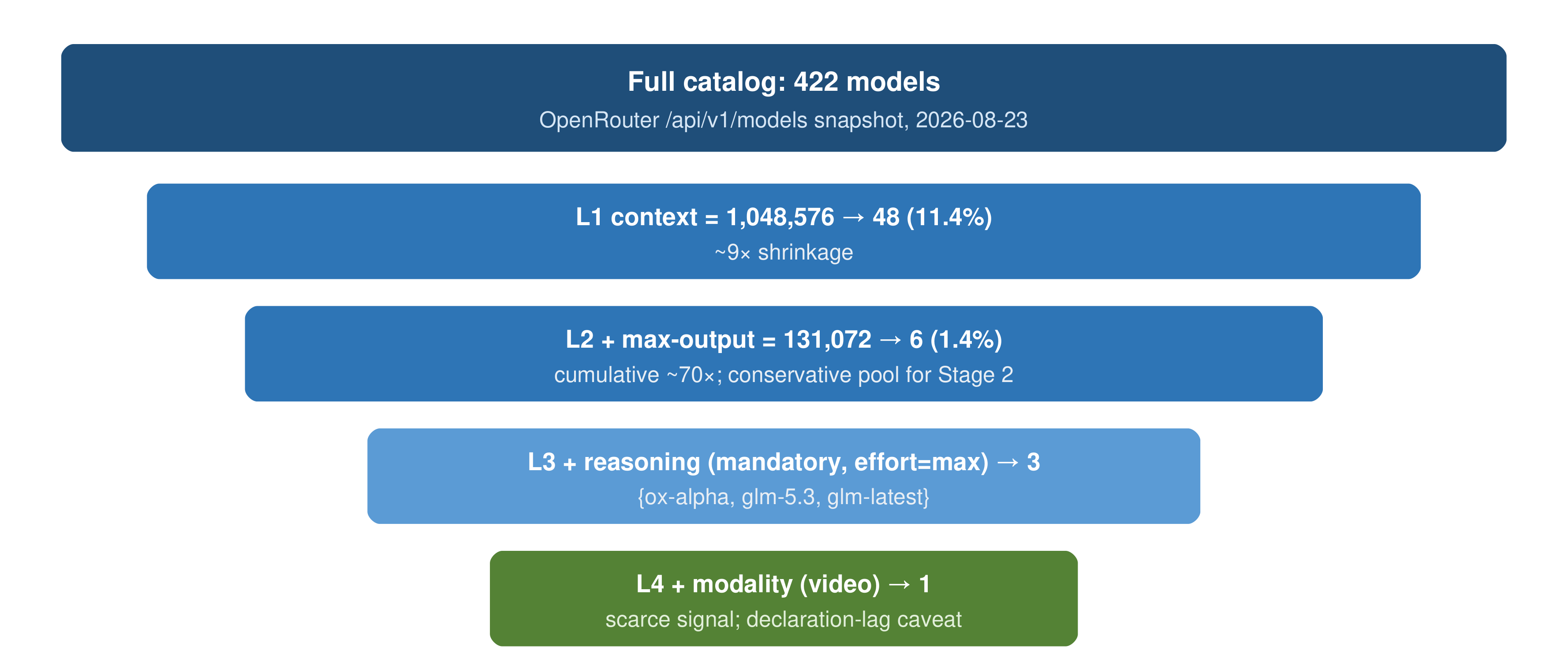}
\caption{Stage 1 layered shrinkage for Ox Alpha against the 2026-08-23 OpenRouter catalog snapshot (422 models). L2 is the conservative Stage-2 pool; L4 modality uniqueness is auxiliary.}
\label{fig:3}
\end{figure}

\subsection{Findings}\label{findings}

\begin{itemize}
\item
  \textbf{F1 (no counter-evidence)}: all 10 cases are same-vendor or compatible: the record-specification channel never pointed to a wrong vendor.
\item
  \textbf{F2 (declaration trustworthiness, revised)}: 9/10 launch-time records matched official specs $\rightarrow$ ``declaration is truthful'' holds \emph{within the same-version semantics}; Polaris is the exception (snapshot 262K vs.~production 400K) $\rightarrow$ declarations drift across versions.
\item
  \textbf{F3 (context inflation is a cross-generation artifact, not same-generation)}: Polaris appears ``inflated'' only against later-generation comparisons (GPT-5.6's 1.05M); against the same-generation GPT-5.1 official 400K it is a snapshot-evolution artifact. Specification signals are valid in launch-time semantics $\rightarrow$ supports Stage 0's time-consistent baseline.
\item
  \textbf{F4 (modality is a scarce time-invariant signal)}: Healer's full modality (text+image+audio+video) matched MiMo-V2-Omni's official multimodal claim; modality is immune to context inflation and version drift, a more robust signal than context length (implemented in Stage 1).
\item
  \textbf{F5 (vendor claim texts are minable primary evidence)}: Xiaomi's official release material explicitly stated ``Hunter Alpha shown below is an early anonymous version of mimo-v2-pro''; vendor official materials are authoritative primary sources for stealth identity (``early anonymous version''/``early snapshot''/``stealth'' as searchable phrases).
\end{itemize}

\subsection{Head-to-head: response-based vs.~configuration-based identification}\label{head-to-head-response-based-vs.-configuration-based-identification}

We test whether deployment properties outperform response behavior on a controlled head-to-head, using LLMmap's released eight-query apparatus and dataset (52 models, mid-2024). We ran the original queries (3 rounds each, 48 queries, zero errors) against GLM-5.3-Flash and GLM-5.3 on the vendor's official endpoint.

On the response channel, single-run outputs were unstable: 14 of 16 model-probe pairs differed across three identical queries. Refusal style, answer format, and self-reported identity did not separate the two versions (both answered as ``GLM by Z.ai''). The flash variant mis-stated its context window as 128K tokens (catalog and Stage 0 baseline: 1,310,720). LLMmap's classifier is trained on 52 mostly $\leq$10B open models plus ten closed APIs, frozen before July 2024, with near-zero Chinese-vendor coverage, outside the 2025--2026 stealth corpus. Self-identification succeeded here only because the endpoints cooperated; the same probes elicited zero credible self-identifications from anonymous Ox Alpha (\textasciitilde300 community attempts).

On the configuration and tokenizer channel, Stage 1 split the versions via reasoning-mandatory flags; Stage 2 separated the family from non-candidates in \textasciitilde10 queries each.

LLMmap's published strength is cross-vendor identification within its training distribution (\textgreater95\% on 42 models). Same-family version discrimination after a model cutoff is a harder task. What this head-to-head shows is narrower than ``configuration beats response'' in general: \textbf{response-style fingerprinting fails on post-cutoff, out-of-distribution stealth models}: instability, cooperative self-ID only, and no version separation, whereas configuration and tokenizer evidence remain usable without retraining. In-distribution LLMmap performance is outside this claim.

\subsection{Negative controls: known backends and non-family collisions}\label{negative-controls-known-backends-and-non-family-collisions}

Beyond the retrospective declaration sample, two control regimes bound false acceptance.

\textbf{(C1) Known-backend re-verification.} After the flagship reveal, the same Stage 2 probes were re-run against the now-public z-ai/glm-5.3-flash OpenRouter endpoint and the vendor-direct API (Section 5.1). Differentials versus the archived anonymous measurements remained constant at 75 across short and long probes on both endpoints, showing that the Stage 2 criterion accepts the true backend when identity is known, and that the anonymous-window fingerprint was not a stealth-only artifact.

\textbf{(C2) Non-family rejection.} The three collision endpoints in Section 3.3 / Fig. 2 / Table 3 are negative controls for the short-prompt failure mode: under a short-only rule they would be accepted; under the mandatory multi-length rule they are rejected. Additional non-GLM candidates in the same archive (e.g., qwen-, ernie-, and mimo-line endpoints in the ox-full JSONL) likewise fail cross-length constancy and are not reported as family matches. Together, C1 and C2 show that Stage 2 is not a one-sided acceptor: it holds for the true family and rejects documented collisions.

\section{Prospective Blind Validation}\label{prospective-blind-validation}

\subsection{Flagship case: Ox Alpha (prediction $\rightarrow$ reveal)}\label{flagship-case-ox-alpha-prediction-reveal}

The protocol was applied to a \emph{live} anonymous model (Ox Alpha, OpenRouter, launched 2026-08-20). On \textbf{2026-08-23}, dated measurement artifacts (catalog snapshot, configuration fingerprint database, and Stage 2 JSONL under supplementary/evidence/) supported an attribution to Zhipu AI (Z.ai), GLM-5.3 version line, on three evidence streams (timeline: Fig. 4). No separate timestamped prediction-deposit file was created that day; the prospective claim is exactly what those artifacts encode. Deployment variant (flash/turbo/max) was \textbf{not} pre-asserted, consistent with contemporaneous community forensics that treated exact serving/quantization/variant as unproven \citep{shrey2026}.

\begin{itemize}
\item
  \textbf{Stage 1 (configuration)}: catalog-wide matching (422 models) reduced the pool to a unique cross-vendor configuration-equivalence class \{ox-alpha, glm-5.3, glm-latest\}, a Zhipu GLM version-line signal.
\item
  \textbf{Stage 2 (tokenizer)}: cross-length differential test over five texts (short$\times$3 + long Chinese + long English): constant differential of 75 tokens vs.~glm-5.3 \emph{and} glm-5.2 across all texts: family-level identity, unique among non-GLM candidates (the three non-family collisions in \S{}3.3 collapsed under the long-text condition). Version-line separation is from Stage 1, not from the tokenizer.
\item
  \textbf{Stage 3 (behavior)}: modality capability (image/video) + reasoning controls matched the GLM serving profile; knowledge-boundary tests suggested an earlier knowledge cutoff (this signal later proved \emph{less} precise than configuration evidence, see below). Modality was recorded as a scarce catalog signal, not as a registered Flash-variant claim.
\end{itemize}

\textbf{Official reveal (2026-08-26)}: Zhipu AI confirmed Ox Alpha = GLM-5.3-Flash (a new-architecture multimodal variant of the GLM-5 generation), released weights, and the stealth entry was delisted. Family and version-line layers matched the 2026-08-23 inferences; Flash as the deployment variant is a \textbf{post-reveal consistency}, not a scored pre-assertion (Table 5):

\begin{table}[H]
\centering
\caption{Prospective flagship case: 2026-08-23 analysis vs official reveal (Ox Alpha = GLM-5.3-Flash).}
\label{tab:5}
\footnotesize
\setlength{\tabcolsep}{4pt}
\begin{tabularx}{\textwidth}{@{}XXXX@{}}
\toprule
Layer & 2026-08-23 analysis & Reveal & Verdict \\
\midrule
Family / generation & Zhipu GLM & GLM-5.3-Flash (GLM-5 gen.) & Confirmed \\
Version line & GLM-5.3 & GLM-5.3-Flash & Confirmed \\
Variant & not pre-asserted & Flash (native multimodal) & Post-reveal consistency only \\
\bottomrule
\end{tabularx}
\end{table}

\begin{figure}[H]
\centering
\includegraphics[width=\textwidth,keepaspectratio]{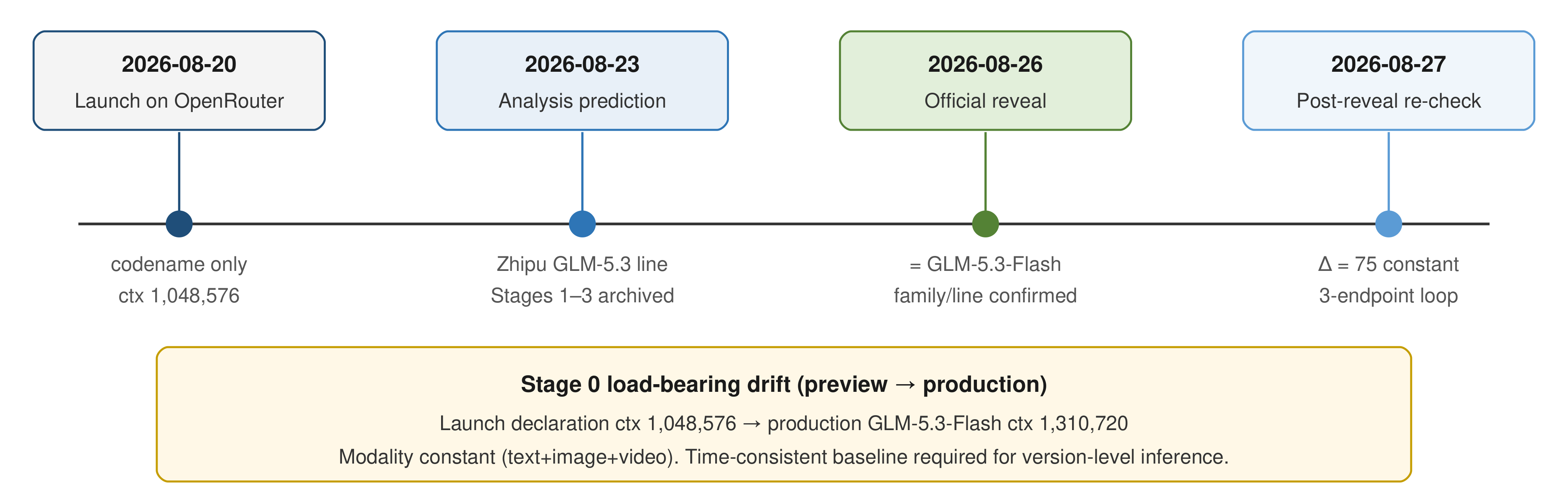}
\caption{Launch $\rightarrow$ 2026-08-23 analysis prediction $\rightarrow$ official reveal $\rightarrow$ post-reveal re-verification, with Stage 0 context drift called out.}
\label{fig:4}
\end{figure}

\textbf{Methodological lessons from the reveal.} (1) \emph{Configuration evidence outperformed knowledge-boundary evidence at version level}: the community forensics campaign's knowledge-boundary inference pointed to the base checkpoint (GLM-5), while our configuration evidence pointed to the version line confirmed by the reveal. We therefore rank knowledge boundaries below configuration and tokenizer evidence (revised Stage 3 ordering). (2) \emph{Tokenizer identity $\neq$ architecture identity}: GLM-5.3-Flash shares the GLM-5-generation tokenizer but is a fully new architecture (linear+sparse attention hybrid, 18B active of 320B total); tokenizer identity established the family, not the checkpoint, consistent with our documented attribution boundary (\S{}3.3). (3) \emph{Version drift between preview and production}: the preview declared 1,048,576 context; the production model declares 1,310,720 (a Polaris-pattern replay). Stage 0's time-consistent baseline is therefore not a formality; it is required for version-level inference. (4) \emph{Community cross-validation}: an independent community forensics campaign (\citealp{shrey2026}; 44-string differentials, 600+ requests) reached the same family-level conclusion through independent methods, and a second independent stylometry study \citep{du2026} likewise favored the GLM-5.3 line while noting it cannot distinguish a base checkpoint from post-training, a system prompt, or a shared serving stack. Both campaigns corroborate the family; our configuration layer provided the finer-grained (version-line) inference, and our tokenizer differential addresses the serving-stack confound those studies self-report.

\textbf{Post-reveal re-verification on the known model.} Four days after the reveal (2026-08-27), we re-ran Stage 1--2 against \emph{two} known endpoints: the now-public z-ai/glm-5.3-flash OpenRouter endpoint and the same model served directly on the vendor's official API, and compared prompt-token counts with the archived pre-reveal measurements of the anonymous model (same probes, same platform, four days apart). The recheck used the four probes that had stable paired counts in the pre-reveal archive under the probe-stability rule (identity\_zh, identity\_en, long\_zh, long\_en); the fifth canonical short probe (cutoff) was not re-paired because the post-reveal flash run did not repeat that wording, so we do not claim a five-of-five recheck. The differential was \textbf{constant at exactly 75 tokens across all four paired probes on both endpoints, including both long-text conditions} (identity\_zh 102$\rightarrow$27/27; identity\_en 102$\rightarrow$27/27; long\_zh 198$\rightarrow$123/123; long\_en 197$\rightarrow$122/122), and every probe returned stable counts across repeated rounds. This closes a three-endpoint loop: anonymous (gateway) $\rightarrow$ public (gateway) $\rightarrow$ official (vendor-direct), with token-identical serving across all three: the fingerprints are stable across the anonymity boundary and are not artifacts of the stealth window or of the gateway's serving configuration. The re-verification also surfaced a contrast at Stage 3: both public endpoints self-report as ``Z.ai GLM'' while the anonymous model suppressed self-identification under identical probing, confirming that self-identification suppression is a deliberate, revocable serving-layer policy rather than a model property (Section 1.2). Full log: supplementary evidence archive.

\subsection{Long-running anonymous case: Big Pickle (two-outcome tracking)}\label{long-running-anonymous-case-big-pickle-two-outcome-tracking}

A second live case, \textbf{Big Pickle} on OpenCode Zen, tests the protocol when ground truth may \emph{never} arrive. Public Zen catalog timelines show a stealth slot occupied by Code Supernova from 2025-10-02, then rotated to Big Pickle from \textbf{2025-11-18}; as of the 2026-08-28 audit window the endpoint had remained anonymous for more than eight months (Supplementary Material S1, cases 19--20).

\textbf{Stage 0--1 case card (archived 2026-08-23; zero API key).} The Zen-facing declaration used for comparison is a \textbf{200K} context window. Against first-party official specifications available at that date:

\begin{table}[H]
\centering
\footnotesize
\setlength{\tabcolsep}{4pt}
\begin{tabularx}{\textwidth}{@{}XXX@{}}
\toprule
Candidate & Official ctx & Match 200K \\
\midrule
GLM-5 & 200K & yes \\
GLM-5.1 & 200K & yes \\
GLM-5.2 & 1M & no (unless Polaris-style drift) \\
Claude-family guesses & varied & inconclusive \\
\bottomrule
\end{tabularx}
\end{table}

\textbf{Archived prediction (deposited 2026-08-24 03:16; target still anonymous; prospective, not post-reveal).} Most-consistent pool: GLM-5 / GLM-5.1 family by configuration exclusion. A GLM-5.2 lead was \textbf{downgraded} on the 1M-vs-200K contradiction (weak without an independent drift archive). Confidence: \textbf{low-to-medium}; Stage 1 narrows; Stage 2--3 were not executed because the Zen channel did not expose usable \emph{prompt\_tokens} usage fields to the auditor's available access (data-availability bound). Full case card: Supplementary Material, Big Pickle prediction note.

\textbf{Two-outcome design.} (i) If later revealed, the archived prediction is scored against the reveal. (ii) If never revealed, the protocol's value is the time-stamped evidence trail itself: candidate pool, excluded lines, confidence ceiling, and the explicit reason Stages 2--3 stopped. This is the operational meaning of \emph{cannot fully attribute} under channel constraints, not a forced guess at a checkpoint.

\subsection{``No-answer'' demonstrations: attribution boundaries}\label{no-answer-demonstrations-attribution-boundaries}

To establish that the protocol declines rather than guesses, we ran Stage-0-only attributions on three delisted cases. Two produced \emph{graded, low-information hypotheses}, not forced identifications: (i) \emph{Cypher} (1M context; GPT-4.1 line hypothesis; medium confidence via exclusion); (ii) \emph{Owl} (1M context; a directional inference later confirmed as LongCat-2.0). The remaining case is a deliberate decline: (iii) \emph{Aurora} (128K context, \emph{cannot attribute}, low confidence, correctly declined because 128K is a common configuration at that time point). No post-hoc evidence contradicted the Aurora decline. We report all three because they show the output discipline, graded hypothesis or cannot-attribute, rather than a closed-set guess.

\section{Portability and Controlled Adversarial Experiments}\label{portability-and-controlled-adversarial-experiments}

The protocol's configuration and tokenizer layers transfer to adjacent audit tasks where ``declared vs.~actual'' divergence is the audit object. Beyond methodological transfer, we ran two \emph{controlled} experiments that exercise the threat-model residual and the serving-switch use case (full logs: Supplementary Material S4; reproducible script under supplementary/experiments/).

\subsection{Experiment E1 --- Usage-forging proxy (A-evasive negative direction)}\label{experiment-e1-usage-forging-proxy-a-evasive-negative-direction}

\textbf{Setup.} Declared family A = GLM-5.3; actual backend B = Kimi. The adversary forges \emph{prompt\_tokens} on every probe to equal A's honest counts (so Stage 2 sees $\Delta$=0 vs an enrolled A reference) while returning B's response text.

\textbf{Offline replay} (archived multi-model JSONL, 2026-08-23): Stage 2 on the forged endpoint is constant at $\Delta$=0 (ACCEPT same-tokenizer as declared). Without forgery, honest B vs A differentials fall in 71--73 (REJECT). Stage 3 identity probes on B self-report as Kimi / Moonshot AI; marker mismatch and identity-content divergence both fire $\rightarrow$ \textbf{cross-stage alarm = true}.

\textbf{Live OpenRouter} (declared z-ai/glm-5.3, backend moonshotai/kimi-k2): same pattern; forged Stage 2 constant $\Delta$=0; honest backend differentials non-constant (range --9\ldots--3); Stage 3 returns Kimi/Moonshot self-identification $\rightarrow$ \textbf{cross-stage alarm = true}.

\textbf{Interpretation.} Asymmetric usage forgery can fool Stage 2 alone; the protocol's defense is the Stage 2--3 inconsistency alarm, not Stage 2 in isolation. Stages 0--2 still defeat non-cooperative \emph{self-identification suppression}; defeating a full-stack usage forger requires retaining Stage 3 as a cross-check.

\subsection{Experiment E2 --- Silent serving switch (downgrade simulation)}\label{experiment-e2-silent-serving-switch-downgrade-simulation}

\textbf{Setup.} An endpoint keeps a fixed declared label. Phase 1 serves premium backend A; phase 2 silently switches to cheaper backend B. The auditor holds an enrolled reference of A and monitors Stage 2 differentials on a fixed probe set.

\textbf{Offline replay} (A=glm-5.3, B=kimi-k3): phase-1 differentials vs reference are constant 0; after switch, differentials jump to 71--73 and lose constancy $\rightarrow$ \textbf{switch detected = true}.

\textbf{Live OpenRouter} (A=z-ai/glm-5.3, B=openai/gpt-4o-mini): phase-1 constant 0 across five probes (including long ZH/EN); phase-2 differentials span --7\ldots23 (non-constant) $\rightarrow$ \textbf{switch detected = true}.

\textbf{Interpretation.} Longitudinal Stage 2 detects backend swaps that change token accounting, without vendor cooperation. This is a controlled simulation of silent downgrade / load-shedding substitution, not a scrape of a commercial ad-funded IDE, so product-specific routing policies remain out of scope. Combined with Section 5.1's anonymity-boundary stability ($\Delta$=75 across anonymous $\rightarrow$ public $\rightarrow$ vendor-direct), Stage 2 is shown to be both \emph{stable when the backend is fixed} and \emph{sensitive when the backend changes}.

\subsection{Branded-model base attribution (methodological transfer)}\label{branded-model-base-attribution-methodological-transfer}

Products increasingly ship with branding over third-party open-weight backbones. Stage 1--2 apply with a changed candidate pool (branded target vs open-weight catalog). We do not claim a branded-product case study here; Experiments E1--E2 and Section 4.5 already show the arithmetic an auditor would reuse.

\section{Discussion}\label{discussion}

\subsection{Relation to disclosure}\label{relation-to-disclosure}

Vendor and platform disclosures remain authoritative when present. The protocol fills the gap before disclosure: it can expose preview--production configuration drift (Stage 0; two cases here) and leave an auditable trail for adoption decisions. On the flagship case, community forensics and our protocol agreed on the family; configuration evidence reached the version line; the vendor reveal settled the remaining layers.

\subsection{Limitations}\label{limitations}

\textbf{Validation as an evidence pyramid, not a single accuracy number.} End-to-end blind identification under anonymity is demonstrated on one flagship case (Ox Alpha; Section 5.1). That case sits atop criterion-level evidence (cross-length collision ablation, Fig. 2), controlled negative checks (E1/E2; Section 6), declaration-channel validation (N=10; Section 4), and cannot-attribute demonstrations, not a substitute for a multi-model prospective panel. We claim a \emph{validated protocol design} with this layered support, not a population identification rate.

Validation covers free-window stealth models only; the paid-stealth cell is empty. Confirmation of identity still depends on eventual disclosure; for never-revealed models the protocol can only emit graded outputs (by design). Platform-reported usage and catalogs can be transformed (Section 2.2); we use differentials rather than absolute counts. Stage 3 is noisier than Stages 0--2 and was ranked downward after the flagship reveal. The prospective identification test is a single comprehensive case; the two-outcome Big Pickle design reduces, but does not remove, single-case dependence. For Ox Alpha, prospective support is encoded in dated measurement artifacts rather than a separately hashed public prediction deposit (Section 5.1); a weaker commitment than the protocol now recommends. Catalogs and terms change, so outputs are time-point valid only.

\subsection{Implications}\label{implications}

For auditors, the protocol is a due-diligence check before adopting an anonymous endpoint and a longitudinal monitor for serving changes. The deliverable is an evidence trail (snapshots, catalog matches, tokenizer differentials, graded confidence), not a classifier score. Where vendors publish Model Cards or platform declarations, Stages 0--1 test whether the served model matches what was declared.

For platforms, anonymity does not hide configuration or tokenizer signals accessible through ordinary API use. Future work should expand platform coverage (Zen endpoints need usable usage fields), fill the paid-stealth cell, and, with institutional access, replicate E2-style switch detection inside production gateways.

\section{Conclusion}\label{conclusion}

Anonymous API endpoints are increasingly released without vendor attribution, yet adopters need auditable identity evidence for supply-chain and data-handling decisions. We specified a four-stage forensic protocol, validated declaration consistency on ten known-identity stealth releases, and prospectively confirmed a flagship case at family and version-line levels after official reveal (deployment variant not pre-asserted). Controlled replays show that forged usage fields trigger a cross-stage alarm and that silent backend switches surface in longitudinal Stage 2. A standard-library audit tool and evidence archive accompany the paper. When disclosure is absent or delayed, measured deployment properties still support due diligence: not certainty, but a traceable audit trail.

\section{Availability}\label{availability}

\begin{itemize}
\item
  \textbf{Tool (audit-tool)}: Python implementation using the standard library only; API keys are read from environment variables and never logged or stored. Source is included in the supplementary archive. The accompanying CLI currently implements Stages 1--2 (catalog fingerprint + cross-length differential) and Stage 3 probes; Stage 0 (Wayback reconstruction) is a documented procedure (Table 2; Supplementary S3), not a packaged crawler.
\item
  \textbf{Case corpus}: the enumerated stealth releases (Table 4 subsample plus remaining cases) are listed as Supplementary Material S1.
\item
  \textbf{Auditor worksheet}: portable forensic output template as Supplementary Material S2.
\item
  \textbf{Stage 0 extracts}: Ox Alpha and Polaris launch-time vs later declarations as Supplementary Material S3.
\item
  \textbf{Controlled experiments}: E1 (usage-forging proxy) and E2 (silent serving switch) logs and script as Supplementary Material S4 (evidence/experiments/, experiments/run\_controlled\_experiments.py).
\item
  \textbf{Evidence archive}: dated 2026-08-23 measurement artifacts for Ox Alpha (catalog snapshot, configuration fingerprint database, Stage 2 JSONL): these encode the prospective GLM-5.3-line attribution; there is no separately hashed public prediction-deposit file for that day (a weaker commitment than the protocol now recommends; Section 3). The Big Pickle case card (\path{evidence/big-pickle-prediction-20260823.md}, deposited 2026-08-24 while the endpoint remained anonymous) is retained as a true prospective record for the never-revealed / long-running design. Raw query logs (credentials removed), tokenizer-differential measurements (Fig. 2 source: \path{evidence/audit-oxfull-20260823-225759.jsonl}; 16-probe archival set with probe-unstable \emph{style\_en} rounds: \path{evidence/probes-16/}), record-page extracts, and post-reveal re-verification logs referenced in Sections 4--5 are included under \path{supplementary/evidence/}. A journal submission may attach the full archive as editorial supplementary material; the public deposit is archived at Zenodo with DOI https://doi.org/10.5281/zenodo.22210928.
\item
  \textbf{Figures}: Fig. 1--4 are provided as vector PDF (Elsevier preferred) plus SVG sources; EM upload copies as Figure\_1.pdf--Figure\_4.pdf; 1000 dpi PNG also available.
\end{itemize}

\section*{CRediT authorship contribution statement}\label{credit-authorship-contribution-statement}

\textbf{Yisen Xi:} Conceptualization; Methodology; Software; Investigation; Data curation; Writing -- original draft; Writing -- review \& editing; Visualization.

\section*{Declaration of competing interest}\label{declaration-of-competing-interest}

The author declares that he has no known competing financial interests or personal relationships that could have appeared to influence the work reported in this paper.

\section*{Funding}\label{funding}

This research did not receive any specific grant from funding agencies in the public, commercial, or not-for-profit sectors.

\section*{Declaration of generative AI and AI-assisted technologies in the manuscript preparation process}\label{declaration-of-generative-ai-and-ai-assisted-technologies-in-the-manuscript-preparation-process}

During the preparation of this work the author used an AI-assisted writing environment (Cursor) for language polishing of the manuscript text. After using this tool, the author reviewed and edited the content as needed and takes full responsibility for the content of the published article. No generative AI tool was used to create or alter figures. No AI tool was used to generate experimental data.

\textbf{Corpus statement.} The phenomenon-level enumeration in the introduction (20 stealth releases, nine vendors, three distribution channels; Supplementary Material S1) was assembled from public sources only: platform record pages, Internet Archive snapshots, vendor release materials, and community documentation. Section 4 tabulates the 10 cases that have \emph{both} a launch-time specification record and a known identity; the remaining enumerated cases lack one of those two conditions and are out of scope for the declaration-consistency sample. No non-public data were used.

\end{document}